\documentclass[
    ,final            
  ]
  {aipproc}

\layoutstyle{6x9}

\begin{document}

\title{Axinos as Cold Dark Matter}


\author{Laura Covi}{
  address={DESY, Notkestrasse 85, D-22603 Hamburg}
}

\begin{abstract}
The connection of Dark Matter (DM) to our particle physics 
model is still one of the open cosmological questions. 
In these proceedings I will argue that axinos
can be successful Cold Dark Matter candidates in models
with Supersymmetry and the Peccei-Quinn solution of the
strong CP problem. If they are the Lightest Supersymmetric
Particle (LSP), they can be produced in the right
abundance by thermal scatterings and out of equilibrium
decays of the Next-to-Lightest Supersymmetric Particles (NLSPs).
Moreover if the NLSPs are charged, their decay could help
us understand which is the LSP, e.g. between axino and
gravitino.
\end{abstract}

\maketitle


\section{Introduction}

The nature of Dark Matter is still unknown today, but from
simulations of structure formation we know it must be a 
neutral, Cold, collision-less (i.e. quite weakly interacting) 
and very long lived particle~\cite{DMreview}. 
Unfortunately such a particle does not exist in the Standard
Model (SM): neutrinos are neutral and massive, but so light that
they are at most Hot Dark Matter. 
Therefore we are obliged to look for DM candidates in models 
beyond the SM.
One of the best motivated extensions, supersymmetry with
R-parity conservation, naturally gives us a stable massive 
particle, the LSP.
To be DM such a particle has to be neutral and very weakly
interacting, so usually only the neutralino or the gravitino
are possible LSPs.
But if we invoke the Peccei-Quinn solution to the strong
CP problem, a new multiplet has to be introduced, 
the axion multiplet. 
As  long as supersymmetry is unbroken, this whole
multiplet remains light, so that no supersymmetric mass 
parameter is allowed for it (contrary than for the Higgses).
After supersymmetry breaking the fermionic component, 
the axino, obtains a mass, but it still could be the LSP and 
make a very good DM component. 
We will present in this talk a summary of axino CDM
\cite{axino1,axino2,axino3,axino4,axino5} and explore
in particular if such particles
can be produced in sufficient numbers to make up most of the
DM and what that implies for the supersymmetry breaking parameters 
and collider searches.

\section{Producing axinos in the Early Universe}

We briefly review here the two main mechanisms that produce 
axinos in the Early Universe. We concentrate here on the 
hadronic type of axion models, where is it expected that 
the axion multiplet does not interact directly with the 
SM multiplets and therefore the axino does not mix 
substantially with the standard neutralinos.
In the other type of models, this mixing can be
larger and the production is therefore enhanced.

\subsection{Thermal scatterings}

Any particle, even very weakly coupled, is produced in
the thermal plasma by scatterings of the particles that are
in thermal equilibrium. Axinos couple directly to the 
gluons and gluinos due to the axion ``anomaly'' coupling
\begin{equation}
W_{PQ} = {g^2 \over 16\sqrt{2} \pi^2 f_a} \Phi_a W^\alpha W_{\alpha}
\quad\rightarrow\quad
{\cal L}_{\tilde a g \tilde g} = {\alpha_s \over 8\pi f_a}
\bar {\tilde a} \gamma_5 \sigma^{\mu\nu} \tilde g^b G^b_{\mu\nu}
\label{dim5op}
\end{equation}
where $\Phi_a $ is the axion multiplet containing the axino $\tilde a$,
$W$ the gluon vector multiplet containing the gluino $\tilde g^b$
and the gluon $G^b_{\mu\nu} $ and $f_a$ is the Peccei-Quinn scale
of the order of $10^{11}$ GeV due to axion physics \cite{axion}.
So many scattering of the primordial plasma involving colored
particles can produce axinos~\footnote{
The same happens also in the case of the gravitino, but with
different vertex structure and scale \cite{gravitino}.}.
The axino number density is given by solving a Boltzmann equation
of the type
\begin{eqnarray}
{d n_{\tilde a} \over d t} + 3 H n_{\tilde a} &=&
\sum_{ij} \langle\sigma (i+j \rightarrow \tilde a + ...) v_{rel} \rangle n_i n_j
+ \sum_{i} \langle\Gamma (i \rightarrow \tilde a + ...)\rangle n_i
\label{Boltzmann}
\end{eqnarray}
where we are neglecting back-reactions, that are suppressed by
$n_{\tilde a} \ll n_i $.

\begin{figure}
  \includegraphics[height=.4\textheight]{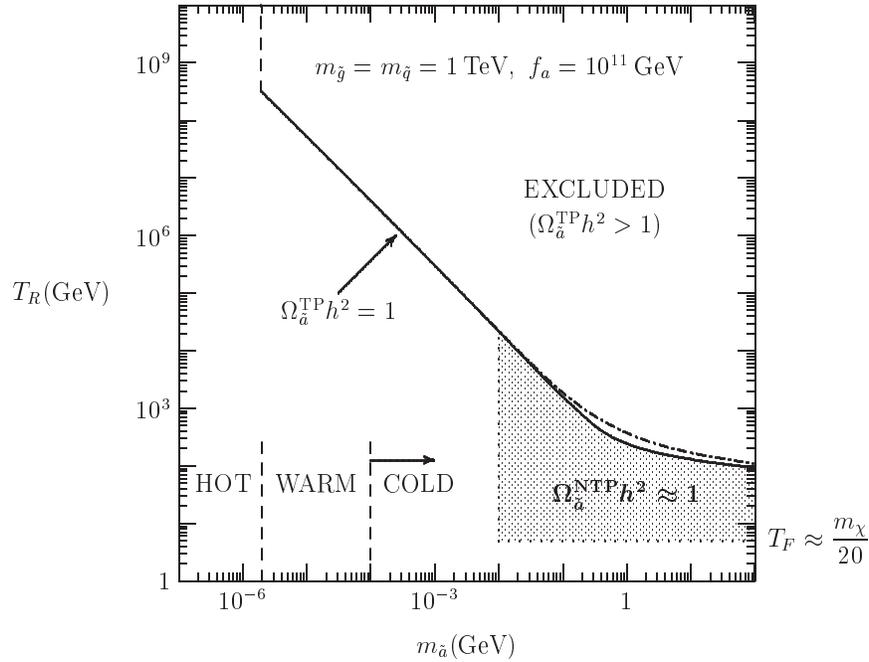}
  \caption{Maximal reheat temperature as a function of the
axino mass obtained by requiring that the axino energy density
is below the present DM density~\cite{axino2}. The difference
between solid and dashed line is due to the inclusion of the
decay term in the Boltzmann equation~(\ref{Boltzmann}).
In the gray area we expect the non-thermal production via
out of equilibrium decays to be also substantial. 
 }
\end{figure}

At high temperature the 2-body scatterings dominate the production, 
since they contain a vertex given by the dimension 5 operator in 
eq.~(\ref{dim5op}) and they show a characteristic linear dependence
on $T$. So most of the axinos are produced at the highest temperature,
the reheat temperature $T_R$, and the axino number density is 
proportional to $T_R$.
Some of the two body scatterings are IR divergent due to the massless 
gluon propagator; in the thermal bath such a divergence is screened 
by the presence of a thermal gluon mass $\simeq g T$.
In our computation we introduced such IR cut-off by hand \cite{axino2}.
A self-consistent procedure is instead to perform a full resummation 
of the Hard Thermal Loops as in \cite{BS04}.

At lower temperatures the decay terms start dominating and the
number density is no more proportional to the reheat temperature,
it depends instead on the supersymmetric parameters, in particular
the gluino and squark masses~\cite{axino3}.

Using the expression for the present axino energy density as
\begin{equation}
m_{\tilde a} {n_{\tilde a} (T)\over s(T)} 
= 0.72\,\mbox{eV}  \left({\Omega_{\tilde a} h^2 \over 0.2 } \right)\; , 
\end{equation}
where $s(T) = 2.89\times 10^3 \left( {T \over 2.726 K} \right)
\mbox{cm}^{-3} $ 
is the present entropy density,
we can then obtain a bound on the reheat temperature as shown in
Figure~1.

\subsection{Out of equilibrium decays}

An axino population is also generated by the NLSP decay after 
it freezes out from the thermal bath.
The heavier superpartners cascade-decay quickly into 
the NLSP (or very rarely  in the LSP as we discussed in 
the previous section) while still in equilibrium,
but the NLSP instead has a lifetime longer than the
freeze-out time: in fact all the axino couplings are 
suppressed by the Peccei-Quinn scale $f_a \simeq 10^{11} $ GeV  
and so the NLSP lifetime is of the order of seconds or longer. 
Then the freeze-out process is unaffected and the decay
takes place only much later as shown in Figure~2.

\begin{figure}
  \includegraphics[height=.4\textheight]{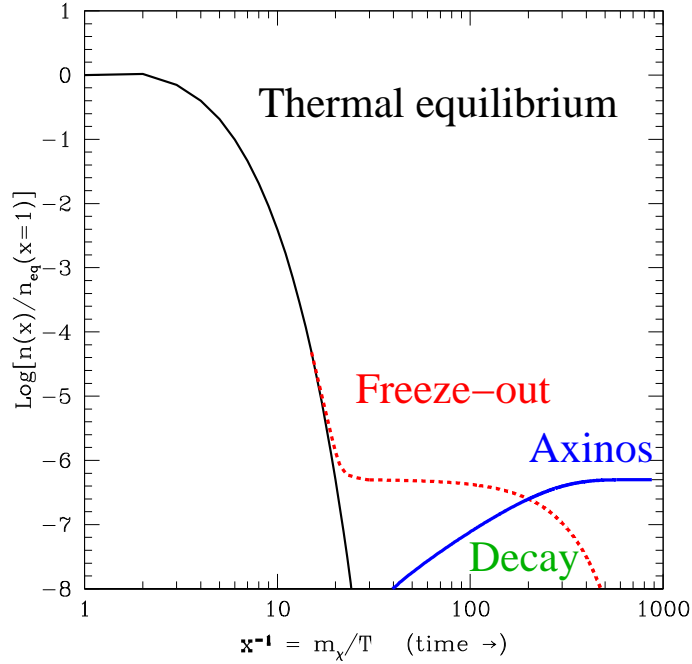}
  \caption{
Freeze-out of the NLSP and subsequent decay into axino.
Due to R-parity conservation the number of axino produced
in the decay is the same as the NLSP number.
}
\end{figure}

In this case, the axino number density can be directly computed 
from the NLSP would-be-relic number density as
\begin{equation}
\Omega_{\tilde a}^{NT} = {m_{\tilde a}\over m_{NLSP}} \; \Omega_{NLSP}.
\label{omegaresc}
\end{equation}
If the mass ratio is not too small, we still have a connection
with the classical WIMP mechanism in case the NLSP is a neutralino 
or a stau.

On the other hand, a couple of problems can arise if the decay
happens too late:

\begin{itemize}

\item
Big Bang Nucleosynthesis can be spoiled by the energetic
``active'' particles produced in the decay along with the axino: 
the strong limits on the injection of energetic particles depend 
on the electromagnetic/hadronic nature of the produced showers 
and the decay time~\cite{BBN}. 
In general such limits are weak for the axino case since the 
NLSP lifetime (excluding a strong mass degeneracy) 
is below $10^2 s $, but they can affect the region of small mass 
for both the neutralino and stau NLSP~\cite{axino2,axino3}.

\item
Are axino from the decay cold enough to be CDM ?
They are relativistic at production even if the NLSP is not
and have a non-thermal spectrum:
\begin{equation}
v (T) = {p(T) \over m_{\tilde a}} \simeq  
{m_{NLSP} \over 2 m_{\tilde a}} 
\left( {g_*(T ) \over g_*(T_{dec})} \right)^{1/3} {T\over T_{dec}},
\end{equation}
where $T_{dec} $ is the temperature at the decay time.
So the question is if they have sufficient time to cool down
before structure formation begins. 
In \cite{JLM05} such constraints have been studied and the
conclusions is that an axino mass at least of order of 1~GeV is
probably needed.

\end{itemize}

\section{Axinos and the CMSSM}

Depending on the parameters and $T_R$, either production mechanism 
can dominate and produce sufficient axinos to explain the present 
DM density.
In general either $T_{R}$ is bounded as in Fig.~1 or the axino 
is so light to be a subdominant (warm or hot) DM component. 
In the last case in our scenario the axion \cite{axion} could be 
the DM.

Assuming that the axinos are CDM and that the supersymmetric 
partners of the SM particles can be described by the Constrained MSSM,
we can see which is the preferred parameter region depending on
the production mechanism.
In the CMSSM the superparticle spectrum and couplings are simply
function of the SM and additional four parameters:
the ratio of the Higgs {\it v.e.v.} $\tan\beta$, 
the gaugino and scalar masses $m_{1/2}, m_0$ and the trilinear 
coupling $A_0$, which are universal at the GUT scale. The modulus 
of the $\mu $ parameter is fixed by radiative electroweak symmetry 
breaking and we will always consider the positive sign in the following.

\subsection{Mostly thermal production}

\begin{figure}
  \includegraphics[height=.4\textheight]{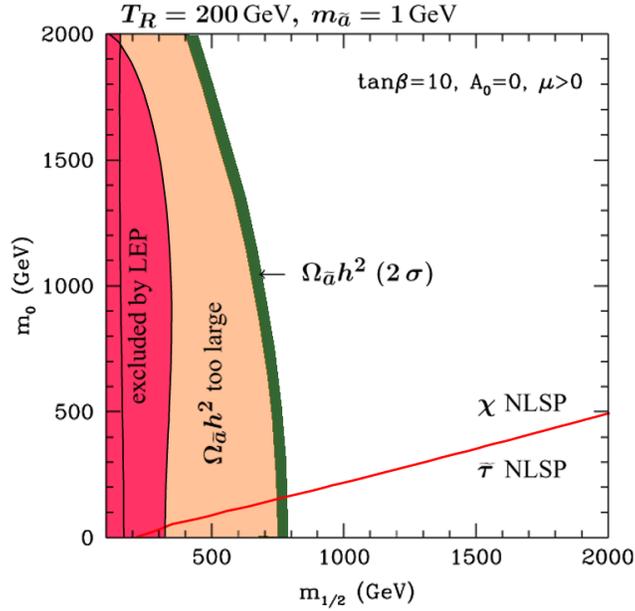}
  \caption{
Allowed parameter space for the case of dominant thermal 
production~\cite{axino4}. We have chosen here $T_R = 200 \mbox{GeV} $,
$m_{\tilde a} = 1$ GeV and $f_a = 10^{11}$ GeV.
The dark gray strip gives axino in the right abundance to explain
all DM, while the lighter gray areas are excluded by LEP constraints
or too large axino number density. The white area has too low
axino density to explain DM.
}
\end{figure}

In the case of high $T_{RH}$ all the particles in the thermal
bath can be treated as massless and so there is practically no 
dependence on the supersymmetry breaking parameters.
On the other hand if we require the axino mass to be above
1 GeV, the reheat temperature has to be sufficiently low and
comparable to the superparter masses, so the decay term in 
the Boltzmann equation become important and a strong dependence 
on the gluino mass appears also due to the squark-quark-axino
coupling~\cite{axino3}.
We have then that the allowed region is a narrow band in the
gaugino mass parameter with a much smaller dependence on $m_0$
as shown in Figure~3.
Note that in this case the small gaugino mass region is excluded
because there too many axinos are produced. The exact position
of the allowed band though strongly depends on the chosen
reheat temperature and it moves at larger $m_{1/2}$ for larger
$T_{R}$.
\vspace*{-0.2cm}

\subsection{Mostly out of equilibrium NLSP decay}

In the CMSSM the NLSP can be either the neutralino or the stau.
The latter happens in the wedge with low $m_0$, that is usually
considered excluded if the stau is the LSP.
In our case though the LSP is the axino and even the stau wedge
is viable; in particular there is a wide region where the stau 
decay can produce the right amount of axino DM as we see in Fig.~4.
\begin{figure}
  \includegraphics[height=.4\textheight]{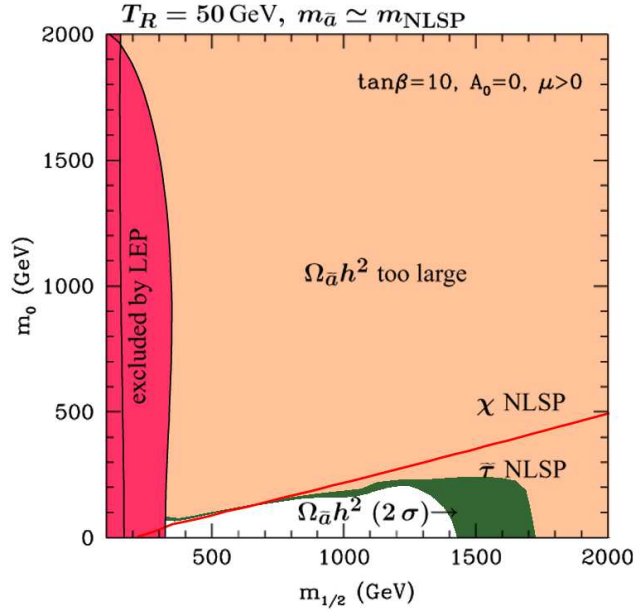}
  \caption{
Allowed parameter space for the case of dominant  
production via out of equilibrium NLSP decay~\cite{axino4}.
We have chosen here $T_R = 200 \mbox{GeV} $,
$m_{\tilde a} = 1$ GeV and $f_a = 10^{11}$ GeV.
The dark gray region gives axino in the right abundance to explain
all DM, while the lighter gray areas are excluded by LEP constraints
or too large axino number density. The white area has too low
axino density to explain DM.
}
\end{figure}
On the other hand there is also a tiny strip in the neutralino
NLSP region, analogous to the neutralino DM region, but shifted
slightly due to the rescaling in eq.~(\ref{omegaresc}).
Both regions are practically unaffected by BBN constraints contrary 
to what happens for the gravitino case~\cite{gravitino2}.
\vspace*{-0.2cm}

\section{How to distinguish the LSP ?}

If the axino is the LSP, very different signals could be found at 
colliders depending on the nature of the NLSP. 
If the neutralino is the NLSP, the only way to find out that it is 
not DM is if the mass and cross sections turn out to give a too 
large neutralino number density or to be excluded by direct DM 
searches. Then we would have good reasons to say that the neutralino
must be unstable, but to study its decay will be very difficult. 

If the stau (or another charged sparticle) is the NLSP instead,
we will have the striking signal of an apparently stable
charged heavy particle in the detector. In that case it will be
clear that the LSP must be a very weakly interacting particle, 
but to know which one, we will need to measure and study the NLSP 
decay.
In particular to distinguish between axino and gravitino, that
can give similar NLSP lifetimes, we will need to measure
the branching ratio and the angular
dependence of the radiative decay in order to reach a
definitive identification~\cite{axino5}.
\vspace*{-0.2cm}

\section{Conclusions}

Axinos with masses in the MeV-GeV range are good CDM candidates
for low reheat~temperature: they can be produced either from
thermal processes or from NLSP decay with the right abundance.
Such scenario is analogous to the gravitino LSP one,
but an axino LSP evades more easily BBN bounds, since the NLSP 
lifetime is shorter than $10^2$ s. Therefore
both neutralino and stau NLSPs are allowed in our case.

Compared to neutralino DM, different regions of the
CMSSM parameter space become allowed and preferred, in particular
even heavier sparticles masses and a charged NLSP like
the stau. An apparently stable charged particle would surely 
give a striking signal at the LHC and ILC and indicate
that the neutralino is not the DM.
On the other hand, disentangling between LSP candidates
will require to stop such NLSPs and measure their decay;
in particular axino and gravitino will be distinguished
if a sufficient number of radiative decay can be observed.
\vspace*{-0.2cm}

\begin{theacknowledgments}
It is a pleasure to thank A. Brandenburg, K. Hamaguchi, 
H.B.~Kim, J.E.~Kim, R. Ruiz de Austri, M. Small, F.D. Steffen
and in particular L. Roszkowski for several years of 
fruitful and exciting collaboration.
The author would also like to thank the Organizers for the 
exciting Workshop and their patience in waiting for these
proceedings.
The author acknowledges the European Network of 
Theoretical Astroparticle Physics ILIAS/N6 under contract 
number RII3-CT-2004-506222 for financial support.
\end{theacknowledgments}





\begin{thebibliography}{9}

\bibitem{DMreview}
  G.~Bertone, D.~Hooper and J.~Silk,
  Phys.\ Rept.\  {\bf 405} (2005) 279
  [arXiv:hep-ph/0404175].


\bibitem{axino1}
 L.~Covi, J.~E.~Kim and L.~Roszkowski,
  Phys.\ Rev.\ Lett.\  {\bf 82} (1999) 4180
  [arXiv:hep-ph/9905212].

\bibitem{axino2}
 L.~Covi, H.~B.~Kim, J.~E.~Kim and L.~Roszkowski,
  JHEP {\bf 0105} (2001) 033
  [arXiv:hep-ph/0101009].

\bibitem{axino3}
  L.~Covi, L.~Roszkowski and M.~Small,
  JHEP {\bf 0207} (2002) 023
  [arXiv:hep-ph/0206119].

\bibitem{axino4}
  L.~Covi, L.~Roszkowski, R.~Ruiz de Austri and M.~Small,
  JHEP {\bf 0406} (2004) 003
  [arXiv:hep-ph/0402240].

\bibitem{axino5}
 A.~Brandenburg, L.~Covi, K.~Hamaguchi, L.~Roszkowski and F.~D.~Steffen,
  Phys.\ Lett.\ B {\bf 617} (2005) 99
  [arXiv:hep-ph/0501287].

\bibitem{axion}
See the contributions of B.~Beltrán and E.~Masso 
to these proceedings.

\bibitem{gravitino}
  M.~Bolz, A.~Brandenburg and W.~Buchmuller,
  Nucl.\ Phys.\ B {\bf 606} (2001) 518
  [arXiv:hep-ph/0012052]. 

\bibitem{BS04}
  A.~Brandenburg and F.~D.~Steffen,
  JCAP {\bf 0408} (2004) 008
  [arXiv:hep-ph/0405158].

\bibitem{BBN}
  M.~Kawasaki, K.~Kohri and T.~Moroi,
  Phys.\ Rev.\ D {\bf 71} (2005) 083502
  [arXiv:astro-ph/0408426].

\bibitem{JLM05}
  K.~Jedamzik, M.~Lemoine and G.~Moultaka,
  JCAP {\bf 0607} (2006) 010
  [arXiv:astro-ph/0508141].


\bibitem{gravitino2}
See the contribution of K.Y. Choi to these proceedings;
also F.~D.~Steffen,
arXiv:hep-ph/0605306 and references therein.

\end{thebibliography}
\end{document}